\documentclass[12pt]{article}
\usepackage{amsmath}
\usepackage{latexsym}
\usepackage{epsfig}
\usepackage{graphics}
\title{\bf On-Shell Calculation of Mixed Electromagnetic and Gravitational Scattering}
\author{Barry R. Holstein\\
Department of Physics-LGRT\\University of Massachusetts\\
Amherst, MA  01003}

\begin{document}
\begin{titlepage}
\maketitle

\begin{abstract}
The study of long-range effects arising from the higher order exchange of massless particles via summation of Feynman diagrams is well known, but recently it has been shown that the use of on-shell methods can provide a streamlined route to the calculation of both electromagnetic and gravitational effects.  In this note we demonstrate that the use of on-shell methods yields a similar simplification in the evaluation of higher order effects in the case of mixed electromagnetic and gravitational scattering.
\end{abstract}
\vfill
\end{titlepage}

\section{Introduction}

At lowest order the Coulomb and Newtonian potentials are well known to arise from single photon or graviton exchange, wherein the transition amplitude involves a propagator having the form $1/q^2$, with $q$ being the four-momentum transfer.  In the nonrelativistic limit, there exists no energy transfer so $1/q^2\rightarrow -1/\boldsymbol{q}^2$, whose Fourier transform yields the characteristic coordinate space $1/r$ behavior, and this derivation has been pointed out by numerous authors.   However, via summation of higher order Feynman diagram contributions, much too has been written about long-range corrections to both electromagnetic and gravitational scattering~\cite{Iw71},\cite {Sp93},\cite{Do94},\cite{Fe88},\cite{Bo03},\cite{Do15},\cite{Kh02}, and it has been demonstrated that the use of on-shell methods can simplify the calculation of each~\cite{Bo14},\cite{Ho17}. Long-range effects have also been evaluated via Feynman diagram methods for mixed electromagnetic and gravitational scattering, which can be thought of as gravitational corrections to electromagnetic scattering~\cite{Ho08},\cite{Bu06},\cite{Fa07},\cite{Bo02}.  The full calculation in this case is a
challenging undertaking, requiring the evaluation of {\it twenty} individual diagrams, with their various signs and statistical factors.  However, it is also possible to evaluate this mixed scattering case via on-shell techniques and that is the purpose of the present note.  As we shall demonstrate, this procedure enormously simplifies the calculation since only {\it two} individual contributions are required.

The basic idea is that in calculating higher order scattering involving massless exchange, as in electromagnetic or gravitational interactions, the leading effects in momentum transfer are of two types.  There exist analytic terms such as polynomials in $t=q^2$ as well as nonanalytic pieces involving $\ln(-t)$, $\sqrt{1\over -t}$, etc.  Fourier transforming from momentum space in order to determine an effective higher order coordinate space potential, the analytic terms yield only local effects.  It is only the nonanalytic pieces which produce long-range forms such as $1/r^n$, $n\ge 2$.  Thus the shape of the long-range effective potential can be determined by identifying the leading nonanalytic terms in loop scattering and this is the technique which has been employed in the work described above.  The conventional method involves constructing the higher order scattering amplitude by addition of the various Feynman diagram contributions to a given process.  However, this procedure can be tedious and requires a careful inclusion of the various signs and statistical factors associated with each diagram.  For this reason the use of on-shell methods has recently been advocated~\cite{Bo14},\cite{Ho17}, the point being that such nonanalytic terms contain t-channel cuts, the discontinuity across which can be determined, using unitarity, by employing {\it on-shell} (physical) amplitudes.  This procedure is very efficient compared to diagrammatic methods since the gauge invariant two photon or two graviton amplitudes used as input already include a Feynman diagram summation and the use of unitarity assures the proper signs and statistical inputs.  A further bonus in the case of gravitation is that there exists no need to include ghost terms, since only {\it physical} amplitudes are required, and the double copy theorem allows graviton amplitudes to be written in terms of their corresponding and much simpler electromagnetic counterparts~\cite{Wh18},\cite{Be10}\cite{Ho06},\cite{Ch95}.  Finally, only two dimensional (solid angle) integration is necessary, rather than the four-dimensional integration involved in Feynman loops.  Such techniques then can lead to a streamlined evaluation of such long-range effects.  Specifically, using the methods developed by Feinberg and Sucher~\cite{Fe88} we have shown how previous diagrammatic results can straightforwardly be obtained for the case of both higher order electromagnetic and gravitational scattering~\cite{Ho17}.  However, since effective long-range potentials have also been calculated for ``mixed" scattering, involving a combination of electromagnetic and gravitational interactions, the purpose of the present note is to show how these same on-shell methods lead also to a simplified evaluation of this case.

Below then in section 2, in order to set notation, we present a brief review of the on-shell method applied to the cases of electromagnetic and gravitational scattering. Then in section 3 we transition to the case at hand---scattering due to the combined effect of electromagnetic and gravitational interactions---which involves an interesting extension of the techniques described in section 2 and is demonstrated to efficiently reproduce previous results obtained by the somewhat more tedious Feynman diagram evaluation. Our conclusions are presented in a brief closing section.

\section{Review Calculation}

In order to set notation for the mixed calculation, we briefly review the use of on-shell methods in the case of higher order electromagnetic and gravitational scattering of two spinless particles~\cite{Fe88},\cite{Ho17}---$A+B\rightarrow \gamma\gamma\rightarrow A+B$ and $A+B\rightarrow gg\rightarrow A+B$.

\subsection{Electromagnetic scattering}

\begin{figure}[ht]
\begin{center}
\epsfig{file=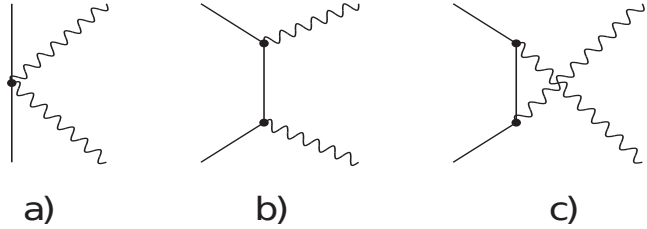,height=3cm,width=8.5cm}
\end{center}
\caption{{Shown are the a) seagull, b) Born, and c) Cross-Born diagrams contributing to Compton scattering. Here the solid lines represent massive
scalars while the wiggly lines are photons.}}\label{comp1}
\end{figure}

We begin with the familiar calculation of Compton scattering---$p_1+\gamma(k_1)\rightarrow p_2+\gamma(k_2)$---from a spinless charged particle of mass $m_A$ and charge $Z_Ae=Z_A\sqrt{4\pi\alpha}$, which arises from the three diagrams in Figure \ref{comp1} and yields
\begin{equation}
{\rm Amp}
^s_{\gamma\gamma}(q)=2Z_A^2e^2\left[\epsilon_2^*\cdot\epsilon_1+{\epsilon_2^*\cdot p_1\epsilon_1\cdot p_2\over p_2\cdot k_1}-{\epsilon_2^*\cdot p_2\epsilon_1\cdot p_1\over p_1\cdot k_1}\right]\label{eq:vv}
\end{equation}

\noindent We next cross to the annihilation channel (t-channel)---$p_1+p_2\rightarrow \gamma(k_1)+\gamma(k_2)$---via the replacement $p_2,k_1\longrightarrow -p_2,-k_1$ in Eq. (\ref{eq:vv})
\begin{equation}
{\rm Amp}^t_{\gamma\gamma}(q)=2Z_A^2e^2\left[\epsilon_2^*\cdot\epsilon_1^*-{\epsilon_2^*\cdot p_1\epsilon_1^*\cdot p_2\over p_2\cdot k_1}-{\epsilon_2^*\cdot p_2\epsilon_1^*\cdot p_1\over p_1\cdot k_1}\right]
\end{equation}
which has the center of mass frame form
\begin{eqnarray}
^{CM}{\rm Amp}^t_{\gamma\gamma}(q)&=&2Z_A^2e^2\hat{\boldsymbol{\epsilon}}_{2j}^*\hat{\boldsymbol{\epsilon}}_{1i}^*\left[-\delta^{ij}+{p^2\hat{\boldsymbol{p}}_A^i\hat{\boldsymbol{p}}_A^j\over E(E+p\cos\hat{\boldsymbol{p}}_A\cdot\hat{\boldsymbol{k}})}+{p^2\hat{\boldsymbol{p}}_A^i\hat{\boldsymbol{p}}_A^j\over E(E-p\cos\hat{\boldsymbol{p}}_A\cdot\hat{\boldsymbol{k}})}\right]\nonumber\\
&=&-2Z_A^2e^2\hat{\boldsymbol{\epsilon}}_{2j}^*\hat{\boldsymbol{\epsilon}}_{1i}^*\left[\delta^{ij}+{2\hat{\boldsymbol{p}}_A^i\hat{\boldsymbol{p}}_A^j\over -{E^2\over p^2}+x_A^2}\right]
\end{eqnarray}
where we have defined $p_1=(E,p\hat{\boldsymbol{p}}_A)$,\,\, $k_1=E(1,\hat{\boldsymbol{k}})$ with $t=4E^2$and have defined $x_A\equiv \hat{\boldsymbol{p}}_A\cdot\hat{\boldsymbol{k}}$.
Following Feinberg and Sucher~\cite{Fe88} we now make the analytic continuation $\boldsymbol{p}_A\rightarrow im_A\xi_A\hat{\boldsymbol{p}}_A$ with $\xi_A=1-{t\over 4m_A^2}$ and define $\tau_A\equiv \sqrt{t}/2m_A\xi_A$.  Then
\begin{equation}
^{CM}{\rm Amp}^t_{\gamma\gamma}\stackrel{\rm anal.\,cont.}{\longrightarrow}-2Z_A^2e^2\hat{\boldsymbol{\epsilon}}_{2j}^*\hat{\boldsymbol{\epsilon}}_{1i}^*{\cal O}_A^{ij}\label{eq:dd}
\end{equation}
with
\begin{equation}
{\cal O}_A^{ij}=\delta^{ij}+{2\over d_A}\hat{\boldsymbol{p}}_A^i\hat{\boldsymbol{p}}_A^j
\end{equation}
where $d_A=\tau_A^2+x_A^2$.  For a pair of spinless particles $A,B$ the unitarity condition for the discontinuity across two photon cut in scalar $A+B\rightarrow A+B$ scattering reads
\begin{eqnarray}
{\rm Disc}\,{\cal M}_{em}(s,t)&=&{-i\over 2!}{(2Z_A^2e^2)(2Z_B^2e^2)\over 4m_Am_B}\int{d^3k_1\over (2\pi)^32k_{10}}{d^3k_2\over (2\pi)^32k_{20}}(2\pi)^4\delta^4(p_1+p_2-k_1-k_2)\nonumber\\
&\times &\sum_{n,m=1}^2{\cal O}_A^{ij}\hat{\boldsymbol{\epsilon}}_{1i}^{n^*}\hat{\boldsymbol{\epsilon}}_{2j}^{m*}\hat{\boldsymbol{\epsilon}}_{1a}^n\hat{\boldsymbol{\epsilon}}_{2b}^m{\cal O}_B^{ab^*}\nonumber\\
&=&-{iZ_A^2Z_B^2e^4\over 16\pi m_Am_B}<{\cal O}_A^{ij}P^T_{ia}(\hat{\boldsymbol{k}})P^T_{jb}(\hat{\boldsymbol{k}}){\cal O}_B^{ab*}>\label{eq:aa}
\end{eqnarray}
where
\begin{equation}
P^T_{ij}(\hat{\boldsymbol{k}})=\sum_{n=1}^2\hat{\boldsymbol{\epsilon}}_i^{n*}\hat{\boldsymbol{\epsilon}}_j^n=\delta_{ij}-\hat{\boldsymbol{k}}_i\hat{\boldsymbol{k}}_j
\end{equation}
is the sum over photon polarizations and $<>$ indicates the solid angle average
\begin{equation}
<G(\hat{\boldsymbol{k}})>\equiv\int{d\Omega_{\hat{\boldsymbol{k}}}\over 4\pi}G(\hat{\boldsymbol{k}})
\end{equation}
where $G(\hat{\boldsymbol{k}})$ is some $\hat{\boldsymbol{k}}$-dependent quantity.
We have divided by the factor $4m_Am_B$ to account for the scalar normalization $<p_2|p_1>=2E\delta^3(\boldsymbol{p}_1-\boldsymbol{p}_2)$.
Performing the polarization contractions indicated in Eq. (\ref{eq:aa}) we find then
\begin{eqnarray}
<{\cal O}_A^{ij}P^T_{ia}(\hat{\boldsymbol{k}})P^T_{jb}(\hat{\boldsymbol{k}}){\cal O}_B^{ab}>&=&<{1\over d_Ad_B}\left[4(y-x_Ax_B)^2-2(1-x_A^2)(1-x_B^2)\right.\nonumber\\
&+&\left.2(1+\tau_A^2)(1+\tau_B^2)\right]>
\end{eqnarray}
where
\begin{equation}
y(s,t)=\hat{\boldsymbol{p}}_A\cdot\hat{\boldsymbol{p}}_B={2s+t-2m_A^2-2m_B^2\over 4m_A\xi_Am_B\xi_B}
\end{equation}
In the small angle scattering limit $t<<s$ and near threshold $s\simeq s_0=(m_A+m_B)^2$ we have then $y,\xi\stackrel{t<<s_0}{\longrightarrow} 1$ and
\begin{eqnarray}
{\rm Disc}\,{\cal M}_{em}(s,t)&\simeq&-{iZ_A^2Z_B^2e_4\over 8\pi m_Am_B}<{1\over d_Ad_B}\left[2-4x_Ax_B+x_A^2+x_B^2+x_A^2x_B^2\right]>\nonumber\\
&=&{-iZ_A^2Z_B^2e^4\over 8\pi m_Am_B}\left(2I_{00}-4I_{11}+I_{20}+I_{02}+I_{22}\right)
\end{eqnarray}
where
\begin{equation}
I_{nm}=<{x_A^nx_B^n\over d_Ad_A}>
\end{equation}
are angular average integrals defined by Feinberg and Sucher, whose values are given, for convenience, in the Appendix.  We find then
\begin{equation}
{\rm Disc}\,{\cal M}_{em}(s,t)=-{iZ_A^2Z_B^2e^4\over 8\pi m_Am_B}\left[{\pi(m_A+m_B)\over \sqrt{t}}+{7\over 3}+i4\pi{m_Am_Bm_r\over p_0t}+\ldots\right]
\end{equation}
where $p_0=\sqrt {m_r(s-s_0)/(m_A+m_B)}$ is the center of mass momentum and $m_r=m_Am_B/(m_A+m_B)$ is the reduced mass.

Since
\begin{equation}
{\rm Disc}\,\left\{\ln(-t),\sqrt{1\over -t}\right\}=\left(2\pi i,-i{2\pi^2\over \sqrt{t}}\right)
\end{equation}
the scattering amplitude itself is
\begin{equation}
{\cal M}_{em}(s,t)=-{Z_A^2Z_B^2\alpha^2\over m_Am_B}\left[-(m_A+m_B)S+{7\over 3}L+4\pi i{m_Am_Bm_r\over p_0t}L+\ldots\right]\label{eq:bb}
\end{equation}
where we have defined $S\equiv\pi^2/\sqrt{-t}$ and $L\equiv\ln(-t)$.  Before attempting to construct an effective potential, it is necessary to remove the
imaginary component of Eq. (\ref{eq:bb}) by subtracting the second order Born iteration of the lowest order Coulomb interaction~\cite{Da51}
\begin{eqnarray}
A^{(2)}_{Born}(q)&=&i\int{d^3\ell\over (2\pi)^3}{Z_AZ_Be^2\over |\boldsymbol{p}_i-\boldsymbol{\ell}|^2+\lambda^2}{i\over {p_i^2\over 2m_r}-{\ell^2\over 2m_r}+i\epsilon}{Z_AZ_Be^2\over |\boldsymbol{p}_f-\boldsymbol{\ell}|^2+\lambda^2}\nonumber\\
&&\stackrel{\lambda\rightarrow 0}{\longrightarrow}-i4\pi Z_A^2Z_B^2\alpha^2{m_r\over p_0t}L\label{eq:ee}
\end{eqnarray}
Finally, the higher order long-range potential can be identified by Fourier transforming the subtracted amplitude
\begin{eqnarray}
V_{eff}^{em}(r)&=&-\int{d^3q\over (2\pi)^3}e^{i\boldsymbol{q}\cdot\boldsymbol{r}}\left({\cal M}_{\gamma\gamma}(q)-A^{(2)}_{Born}(q)\right)\nonumber\\
&=&-{Z_A^2Z_B^2\alpha^2(m_A+m_B)\over 2\pi^2r^2}-{7Z_A^2Z_B^2\alpha^2\hbar\over 6\pi^3r^3}
\end{eqnarray}
which is the form we were seeking and is in agreement with the result found by Feynman diagram methods in~\cite{Fe88} and \cite{Hr08}.

\subsection{Gravitational Scattering}

\begin{figure}[ht]
\begin{center}
\epsfig{file=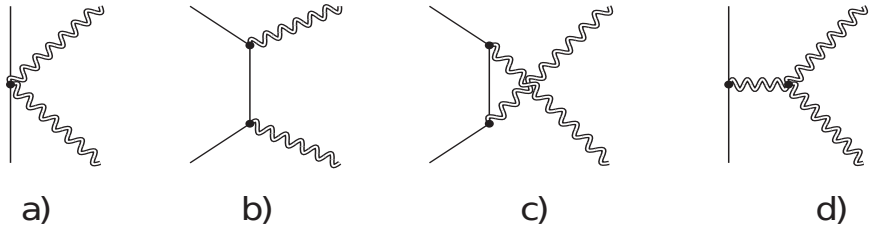,height=3cm,width=11.5cm} \caption{Contact a), Born b),c) and graviton pole d) diagrams
relevant for gravitational Compton scattering. Here the double wiggly lines represent gravitons while the solid lines are massive scalars.}\label{comp3}
\end{center}
\end{figure}

For completeness, and since it introduces two important new ingredients, we also briefly summarize the corresponding gravitational scattering case, for which the conventional Feynman diagram calculation can be found in \cite{Bo03} and \cite{Kh02}.  The corresponding gravitational on-shell calculation is given in \cite{Bo14} and \cite{Ho17}.  The gravitational Compton amplitude--$-p_1+g(k_1)\rightarrow p_2+g(k_2)$---arises from the four diagrams shown in Figure \ref{comp3} but can be enormously simplified by use of the double copy theorem, which allows the gravitational Compton amplitude to be written in a factorized form as a product of ordinary Compton amplitudes multiplied by a simple kinematic factor~\cite{Wh18},\cite{Be10}\cite{Ho06},\cite{Ch95}.
\begin{equation}
{\rm Amp}^s_{gg}(q)={\kappa^2\over 8Z_A^4e^4}K({\rm Amp}^s_{\gamma\gamma}(q))^2
\end{equation}
where $\kappa=\sqrt{32\pi G}$ is the gravitational coupling and
\begin{equation}
K={p_1\cdot k_1p_2\cdot k_1\over k_1\cdot k_2}
\end{equation}
is a kinematic factor.

What is required for the discontinuity calculation is the t-channel (annihilation) amplitude---$p_1+p_2\rightarrow g(k_1)+g(k_2)$.  In the t-channel center of mass frame we have then
\begin{equation}
^{CM}K^t={E(E-px_A)E(E+px_A)\over 2E^2}={p^2\over 2}\left({E^2\over p^2}-x_A^2\right)\stackrel{\rm anal.\,cont.}{\longrightarrow}{1\over 2}m_A^2\xi_A^2d_A
\end{equation}
and can write\footnote{Note that we are working in deDonder gauge, wherein the graviton polarization tensor can be written as a simple product of two vector polarizations.}
\begin{equation}
^{CM}{\rm Amp}^t_{gg}(q)\stackrel{\rm anal.\,cont.}{\longrightarrow}{\kappa^2m_A^2\xi_A^2d_A\over 4}\hat{\boldsymbol{\epsilon}}_{2r}^*\hat{\boldsymbol{\epsilon}}_{2u}^*\hat{\boldsymbol{\epsilon}}_{1s}^*\hat{\boldsymbol{\epsilon}}_{1v}^*{\cal O}_A^{rs}{\cal O}_A^{uv}
\end{equation}
so
\begin{eqnarray}
{\rm Disc}\,{\cal M}_{grav}(s,t)&=&{-i\over 2!}{(\kappa^2m_A^2\xi_A^2)(\kappa^2m_B^2\xi_B^2)\over 64m_Am_B}\int{d^3k_1\over (2\pi)^32k_{10}}{d^3k_2\over (2\pi)^32k_{20}}(2\pi)^4\delta^4(p_1+p_2-k_1-k_2)\nonumber\\
&\times&d_Ad_B\sum_{n,m=1}^2{\cal O}_A^{ij}{\cal O}_A^{k\ell}\hat{\boldsymbol{\epsilon}}_{1i}^{n^*}\hat{\boldsymbol{\epsilon}}_{2j}^{m*}\hat{\boldsymbol{\epsilon}}_{1k}^{n^*}
\hat{\boldsymbol{\epsilon}}_{2\ell}^{m^*}\hat{\boldsymbol{\epsilon}}_{1a}^n\hat{\boldsymbol{\epsilon}}_{2b}^m
\hat{\boldsymbol{\epsilon}}_{1c}^n\hat{\boldsymbol{\epsilon}}_{2d}^m{\cal O}_B^{ab^*}{\cal O}_B^{cd^*}\nonumber\\
&=&-i{\kappa^4m_A^2\xi_A^2m_B^2\xi_B^2\over 1024\pi m_Am_B}<d_Ad_B{\cal O}_A^{ij}{\cal O}_A^{k\ell}Q^T_{ik;ac}(\hat{\boldsymbol{k}})Q^T_{j\ell;bd}(\hat{\boldsymbol{k}}){\cal O}_B^{ab*}{\cal O}_B^{cd*}>\label{eq:rr}
\end{eqnarray}
where
\begin{equation}
Q^T_{jk;bc}(\hat{\boldsymbol{k}})=\sum_{n=1}^2\hat{\boldsymbol{\epsilon}}_{j}^{m*}\hat{\boldsymbol{\epsilon}}_{k}^{n*}
\hat{\boldsymbol{\epsilon}}_{b}^m\hat{\boldsymbol{\epsilon}}_{c}^n={1\over 2}\left(
P^T_{jb}(\hat{\boldsymbol{k}})P^T_{kc}(\hat{\boldsymbol{k}})+P^T_{jc}(\hat{\boldsymbol{k}})P^T_{kb}(\hat{\boldsymbol{k}})
-P^T_{jk}(\hat{\boldsymbol{k}})P^T_{bc}(\hat{\boldsymbol{k}})\right)
\end{equation}
is the sum over graviton polarizations.  Performing the polarization contractions in Eq. (\ref{eq:rr}) we find
\begin{eqnarray}
&&<d_Ad_B{\cal O}_A^{ij}{\cal O}_A^{k\ell}Q^T_{ik;ac}(\hat{\boldsymbol{k}})Q^T_{j\ell;bd}(\hat{\boldsymbol{k}}){\cal O}_B^{ab*}{\cal O}_B^{cd*}>
\stackrel{t<<m_A^2m_B^2}{\longrightarrow}<{2\over d_Ad_B}\left[8(1-x_Ax_B)^4\right.\nonumber\\
&-&\left.8(1-x_Ax_B)^2(1-x_A^2)(1-x_B^2)+(1-x_A^2)^2(1-x_B^2)^2+1\right]>
\end{eqnarray}
Then
\begin{eqnarray}
{\rm Disc}\,{\cal M}_{grav}(s,t)&=&-i{\kappa^4m_Am_Am_B\over 512\pi}\left[2I_{00}-16I_{11}+6I_{20}+6I_{02}+36I_{22}+I_{40}+I_{04}\right.\nonumber\\
&-&\left.16I_{31}-16I_{13}-16I_{33}+6I_{42}+6I_{24}+I_{44}\right]\nonumber\\
&=&-i{\kappa^4m_Am_B\over 512\pi}\left[6{\pi(m_A+m_B)\over \sqrt{t}}+{41\over 5}+i4\pi{m_Am_Bm_r\over p_0t}+\ldots\right]
\end{eqnarray}
so that the gravitational scattering amplitude is
\begin{equation}
{\cal M}_{grav}(s,t)={\kappa^4m_Am_B\over 1024\pi^2}\left[6S(m_A+m_B)-{41\over 5}L-4\pi i{m_Am_Bm_r\over p_0t}L+\dots\right]
\end{equation}
As in the electromagnetic case there exists a second order Born Amplitude
\begin{eqnarray}
B^{(2)}_{Born}(q)&=&i\int{d^3\ell\over (2\pi)^3}{{1\over 8}\kappa^2m_A^2\over |\boldsymbol{p}_i-\boldsymbol{\ell}|^2+\lambda^2}{i\over {p_i^2\over 2m_r}-{\ell^2\over 2m_r}+i\epsilon}{{1\over 8}\kappa^2m_B^2\over |\boldsymbol{p}_f-\boldsymbol{\ell}|^2+\lambda^2}\nonumber\\
&&\stackrel{\lambda\rightarrow 0}{\longrightarrow}-i4\pi G^2m_A^2m_B^2{m_r\over p_0t}L\label{eq:ee}
\end{eqnarray}
which must be subtracted, yielding the effective potential
\begin{eqnarray}
V_{eff}^{grav}(r)&=&-\int{d^3q\over (2\pi)^3}e^{-i\boldsymbol{q}\cdot\boldsymbol{r}}\left({\cal M}_{gg}(q)-B^{(2)}_{Born}(q)\right)\nonumber\\
&=&-{3G^2m_Am_B(m_A+m_B)\over r^2}-{41G^2m_Am_B\hbar\over 10\pi r^3}
\end{eqnarray}
which agrees with the result found via Feynman diagrams~\cite{Bo03},\cite{Kh02}.  Our goal now is to extend these electromagnetic and gravitational calculations to the case of mixed electromagnetic-gravitational scattering.

\section{Mixed Calculation}

\begin{figure}[ht]
\begin{center}
\epsfig{file=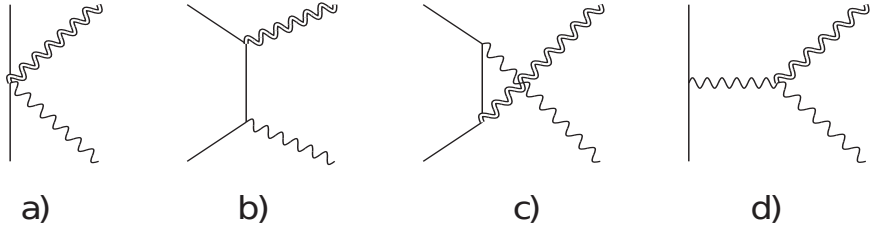,height=3cm,width=11.5cm} \caption{Contact a), Born b),c) and photon pole d) diagrams
relevant for graviton photoproduction. Here the single wiggly lines represent photons, the double wiggly lines represent gravitons while the solid lines
are massive scalars.}\label{comp2}
\end{center}
\end{figure}

The calculation of scalar scattering $A+B\rightarrow A+B$ to ${\cal O}(G\alpha)$ involves important changes from the purely electromagnetic ${\cal O}(\alpha^2)$ and purely gravitational ${\cal O}(G^2)$ cases described in the previous section.  One is that there are now twenty independent Feynman diagrams which contribute rather than the five (eleven) which are relevant in the purely electromagnetic (gravitational) cases. The specific figures can be found in \cite{Bo02} and we shall refer to this reference rather than draw the many individual diagrams herein. A second challenge is to include graviton- in addition to photon-propagation while a third difference is that there are now {\it two} different sources of t-channel cut discontinuities, one from the graviton-photon intermediate state and another from a two photon contribution.  In order to evaluate the former we begin with the amplitude for graviton photoproduction, which arises from the four diagrams shown in Figure \ref{comp2}. However, the calculation is enormously simplified by use of the double copy theorem~\cite{Wh18},\cite{Be10},\cite{Ho06},\cite{Ch95} which asserts that the graviton photoproduction amplitude---$p_1+\gamma(k_1)\rightarrow p_2+g(k_2)$---can be written in terms of the corresponding Compton scattering amplitude by
\begin{equation}
{\rm Amp}^s_{g\gamma}(q)={\kappa\over 2e}H^s{\rm Amp}^s_{\gamma\gamma}(q)
\end{equation}
where $H^s$ is the factor
\begin{equation}
H^s={\epsilon_2^*\cdot p_2k_2\cdot p_1-\epsilon_2^*\cdot p_1k_2\cdot p_2\over k_1\cdot k_2}
\end{equation}
For the discontinuity calculation we require the t-channel (annihilation) amplitude---$p_1+p_2\rightarrow\gamma(k_1)+g(k_2)$---and is obtained via the crossing transformation $p_2,k_1\rightarrow -p_2,-k_1$ as before.  In the t-channel center of mass frame then
\begin{equation}
^{CM}H^t={\hat{\boldsymbol{\epsilon}}_2^*\cdot\hat{\boldsymbol{p}}_ApE(E+px_A)+\hat{\boldsymbol{\epsilon}}_2^*\cdot\hat{\boldsymbol{p}}_ApE(E-px_A)\over 2E^2}=p\hat{\boldsymbol{\epsilon}}_2^*\cdot\hat{\boldsymbol{p}}_A\stackrel{\rm anal.\,cont.}{\longrightarrow}im_A\xi_A\hat{\boldsymbol{\epsilon}}_2^*\cdot\hat{\boldsymbol{p}}_A
\end{equation}

\noindent and the center of mass annihilation amplitude---$p_1+p_2\rightarrow\gamma(k_1)+g(k_2)$---takes the form
\begin{equation}
^{CM}{\rm Amp}^t_{g\gamma}(q)\stackrel{\rm anal.\,cont.}{\longrightarrow}-\kappa Z_Ae m_A\xi_A\hat{\boldsymbol{\epsilon}}_{2k}^*\hat{\boldsymbol{\epsilon}}_{2j}^*\hat{\boldsymbol{\epsilon}}_{1i}^*{\cal U}_A^{jk;i*}
\end{equation}
where
\begin{equation}
{\cal U}_A^{i;jk}={\cal O}_A^{ij}\hat{\boldsymbol{p}}_A^k=\left(\delta^{ij}+{2\over d_A}\hat{\boldsymbol{p}}_A^i\hat{\boldsymbol{p}}_A^j\right)\hat{\boldsymbol{p}}_A^k
\end{equation}
We note that fourteen of the twenty diagrams contributing to the mixed scattering process involve t-channel $g\gamma$ exchange---$p_{A1}+p_{A2}\rightarrow\gamma(k_1)+g(k_2)\rightarrow p_{1B}+p_{2B}$---Figures 2,3,4,5,8 in \cite{Bo02}.  We can then write the discontinuity of the $A-B$ scalar mixed scattering amplitude across the $g\gamma$ cut as\footnote{Note that, unlike the $gg$ and $\gamma\gamma$ case there is no factor ${1\over 2!}$ here since the exchanged massless particles are not identical.}
\begin{eqnarray}
{\rm Disc}\,{\cal M}_{grav-em}^{g\gamma}(s,t)&=&i\left(\kappa Z_Ae\xi_A\right)\left(\kappa Z_Be\xi_B\right)\int{d^3k_1\over (2\pi)^32k_{10}}{d^3k_2\over (2\pi)^32k_{20}}(2\pi)^4\delta^4(p_1+p_2-k_1-k_2)\nonumber\\
&\times &\sum_{n,m=1}^2{\cal U}_A^{i;jk}\hat{\boldsymbol{\epsilon}}_{1i}^{n^*}\hat{\boldsymbol{\epsilon}}_{2j}^{m*}
\hat{\boldsymbol{\epsilon}}_{2k}^{m*}\hat{\boldsymbol{\epsilon}}_{1a}^n\hat{\boldsymbol{\epsilon}}_{2b}^m\hat{\boldsymbol{\epsilon}}_{2c}^m{\cal U}_B^{a;bc*}\nonumber\\
&=&-{i\kappa^2Z_AZ_Be^2\xi_A\xi_B\over 32\pi}<{\cal U}_A^{i;jk}P^T_{ia}(\hat{\boldsymbol{k}})Q^T_{jk;bc}(\hat{\boldsymbol{k}}){\cal U}_B^{a;bc}>\label{eq:ff}
\end{eqnarray}
Performing the polarization contractions in Eq. (\ref{eq:ff}) we find
\begin{eqnarray}
&&<{\cal U}_A^{i;jk}P^T_{ia}(\hat{\boldsymbol{k}})Q^T_{jk;bc}(\hat{\boldsymbol{k}}){\cal U}_B^{a;bc *}>=<{1\over d_Ad_B}\left[8(y-x_Ax_B)^2-4(1-x_A^2)(1-x_B^2)\right.\nonumber\\
&&\left.+2d_A(1-x_B^2)+2d_B(1-x_A^2)+d_Ad_B\right]>\nonumber\\
&&\stackrel{t<<m_A^2m_B^2}{\longrightarrow}<{1\over d_Ad_B}\left[2-10x_Ax_B+3x_A^2+3x_B^2+9x_A^2x_B^2-3x_A^3x_B-3x_Ax_B^3-x_A^3x_B^3\right]>\nonumber\\
&=&2I_{00}-10I_{11}+3I_{20}+3I_{02}+9I_{22}-3I_{31}-3I_{13}-I_{33}\nonumber\\
&=&3{\pi(m_A+m_B)\over \sqrt{t}}+6+i4\pi{m_Am_Bm_r\over p_0t}
\end{eqnarray}
We have then for the $g\gamma$ cut discontinuity
\begin{equation}
{\rm Disc}\,{\cal M}^{g\gamma}_{grav-em}(s,t)=-{i\kappa^2Z_AZ_Be^2\over 32\pi}\left[3{\pi(m_A+m_B)\over \sqrt{t}}+6+i4\pi{m_Am_Bm_r\over p_0t}\right]
\end{equation}

However, unlike the electromagnetic and gravitational cases there exists an additional contribution arising the six diagrams having t-channel two-photon exchange associated with the contraction of an electromagnetic Compton annihilation amplitude, say for particle B, with that for particle A generated by the graviton pole diagram---$p_{A1}+p_{A2}\rightarrow g\rightarrow \gamma(k_1)+\gamma(k_2)\rightarrow p_{B1}+p_{B2}$---Figures 6 and 7 in \cite{Bo02}.  Of course, there exist two categories of $\gamma\gamma$ cut contributions here, since the graviton pole diagram can be associated with either particle $A$ or particle $B$.  First suppose that the graviton pole Compton amplitude is connected with particle A and the electromagnetic Compton amplitude with particle B. Writing $g_{\mu\nu}=\eta_{\mu\nu}+\kappa h_{\mu\nu}$ the gravitational interaction is generated by the vertex
\begin{equation}
{\cal L}_{int}={\kappa\over 2}T_{\mu\nu}h^{\mu\nu}
\end{equation}
where
\begin{equation}
T_{\mu\nu}=2{\delta{\cal L}_{mat}\over \delta g^{\mu\nu}}-g_{\mu\nu}{\cal L}_{mat}
\end{equation}
is the energy-momentum tensor.  Then for the scalar field we have
\begin{equation}
{\cal L}_{mat}={1\over 2}\partial_\mu\phi g^{\mu\nu}\partial_{\nu}\phi-{1\over 2}m^2\phi^2
\end{equation}
so
\begin{equation}
T_{\mu\nu}=\partial_\mu\phi\partial_\nu\phi-{1\over 2}g_{\mu\nu}\left(\partial_\mu\phi g^{\mu\nu}\partial_\nu\phi-m^2\phi^2\right)
\end{equation}
and the scalar vertex becomes
\begin{equation}
\tau^{(0)}_{\mu\nu}(p_1,p_2)={\kappa\over 2}<p_2|T_{\mu\nu}|p_1>={\kappa\over 2}\left[p_{1\mu}p_{2\nu}+p_{1\nu}p_{2\mu}-\eta_{\mu\nu}(p_1\cdot p_2-m^2)\right]\label{eq:zz}
\end{equation}
For the photon we have
\begin{equation}
{\cal L}_{mat}=-{1\over 4}F_{\mu\alpha}g^{\mu\nu}g^{\alpha\beta} F_{\nu\beta}
\end{equation}
so
\begin{equation}
T_{\mu\nu}=F_{\mu\alpha}F_\nu^\alpha-{1\over 4}g_{\mu\nu}F_{\alpha\beta}F^{\alpha\beta}
\end{equation}
and the photon vertex is
\begin{eqnarray}
\epsilon_2^{\beta *}\epsilon_1^\alpha {^\gamma} \tau^{(\gamma)}_{\alpha\beta;\mu\nu}(k_1,k_2)&=&{\kappa\over 2}<k_2,\epsilon_2|T_{\mu\nu}|k_1,\epsilon_1>\nonumber\\
&=&{\kappa\over 2}\epsilon_2^{\beta *}\epsilon_1^\alpha\left[
\left(\eta_{\alpha\mu}\eta_{\beta\nu}+\eta_{\alpha\nu}\eta_{\beta\mu}-\eta_{\alpha\beta}\eta_{\mu\nu}\right)k_1\cdot k_2\right.\nonumber\\
&+&\left.\eta_{\mu\nu}k_{1\beta}k_{2\alpha}+\eta_{\alpha\beta}\left(k_{1\mu}k_{2\nu}+k_{1\nu}k_{2\mu}\right)\right.\nonumber\\
&-&\left.k_{2\alpha}\left(k_{1\mu}\eta_{\beta\nu}+k_{1\nu}\eta_{\beta\mu}\right)\right.\nonumber\\
&-&\left.k_{1\beta}\left(k_{2\mu}\eta_{\alpha\nu}+k_{2\nu}\eta_{\alpha\mu}\right)\right]\label{eq:vv}
\end{eqnarray}
Contraction with the graviton propagator
\begin{equation}
D_{\alpha\beta;\gamma\delta}(q)={i\over 2q^2}\left(\eta_{\alpha\gamma}\eta_{\beta\delta}+\eta_{\alpha\delta}\eta_{\beta\gamma}
-\eta_{\alpha}\eta_{\gamma\delta}\right)
\end{equation}
then yields the graviton pole Compton amplitude
\begin{eqnarray}
{\rm Amp}^s_{g\gamma}(q)&=&\left({\kappa\over 2}\right)^2\epsilon_2^{\beta *}{\epsilon_1^\alpha} ^\gamma\tau^{(\gamma)}_{\alpha\beta;\mu\nu}(k_1,k_2)D^{\mu;\rho\sigma}(q) ^0\tau^{(0)}_{\rho\sigma}(p_1,p_2)\nonumber\\
&=&\epsilon_2^{\beta *}\epsilon_1^\alpha {\kappa^2\over 4q^2}\left[k_1\cdot k_2\left(p_{1\alpha}p_{2\beta}
+p_{1\beta}p_{2\alpha}-\eta_{\alpha\beta}p_1\cdot p_2\right)\right.
\nonumber\\
&+&\left.p_1\cdot p_2k_{1\beta}k_{2\alpha}+\eta_{\alpha\beta}\left(p_1\cdot k_1p_2\cdot k_2+p_1\cdot k_2p_2\cdot k_1\right)\right.\nonumber\\
&-&\left.k_{2\alpha}\left(p_1\cdot k_1p_{2\beta}+p_2\cdot k_1p_{1\beta}\right)\right.\nonumber\\
&-&\left.k_{1\beta}\left(p_1\cdot k_2p_{2\alpha}+p_2\cdot k_2p_{1\alpha}\right)\right]
\end{eqnarray}
After crossing the t-channel form is unchanged and in the center of mass frame becomes
\begin{eqnarray}
^{CM}{\rm Amp}^t_{g\gamma}(q)&=&{\kappa^2\over 16E^2}\hat{\boldsymbol{\epsilon}}_{1i}^*\hat{\boldsymbol{\epsilon}}_{2j}^*\left\{2E^2\left[\delta^{ij}
(E^2+p^2)-2p^2\hat{\boldsymbol{p}}_A^j\hat{\boldsymbol{p}}_A^j\right]\right.\nonumber\\
&-&\left.\delta^{ij}E^2\left[
(E-px_A)^2+(E+px_A)^2\right]\right\}\nonumber\\
&=&{\kappa^2p^2\over 8}\hat{\boldsymbol{\epsilon}}_{1i}^*\hat{\boldsymbol{\epsilon}}_{2j}^*\left[\delta^{ij}(1-x_A^2)-2
\hat{\boldsymbol{p}}_A^j\hat{\boldsymbol{p}}_A^j\right]\stackrel{\rm anal.\, cont.}{\longrightarrow}-{\kappa^2\over 8}m_A^2\xi_A^2\hat{\boldsymbol{\epsilon}}_{1i}^*\hat{\boldsymbol{\epsilon}}_{2j}^*{\cal V}^{ij}\nonumber\\
\qquad\label{eq:cc}
\end{eqnarray}
with
\begin{equation}
{\cal V}^{ij}=\delta^{ij}(1-x_A^2)-2\hat{\boldsymbol{p}}_A^j\hat{\boldsymbol{p}}_A^j
\end{equation}
Combining Eq. (\ref{eq:cc}) with the electromagnetic Compton amplitude Eq. (\ref{eq:dd}) associated with particle $B$ we determine the discontinuity across the two photon cut to be
\begin{eqnarray}
{\rm Disc}\,{\cal M}_{grav-em}^{\gamma\gamma}(s,t)&=&{-i\over 2!}{({\kappa^2m_A^2\xi_A^2\over 8})(2Z_B^2e^2)\over 4m_Am_B}\int{d^3k_1\over (2\pi)^32k_{10}}{d^3k_2\over (2\pi)^32k_{20}}(2\pi)^4\delta^4(p_1+p_2-k_1-k_2)\nonumber\\
&\times &\sum_{n,m=1}^2{\cal V}_A^{ij}\hat{\boldsymbol{\epsilon}}_{1i}^{n^*}\hat{\boldsymbol{\epsilon}}_{2j*}^{m}\hat{\boldsymbol{\epsilon}}_{1a}^n\hat{\boldsymbol{\epsilon}}_{2b}^m{\cal O}_B^{ab^*}+B\leftrightarrow A\nonumber\\
&=&-i{\kappa^2Z_B^2e^2\over 64\pi}{m_A\over m_B}<{\cal V}_A^{ij}P^T_{ia}(\hat{\boldsymbol{k}})P^T_{jb}(\hat{\boldsymbol{k}}){\cal O}^{ab*}>
+B\leftrightarrow A\label{eq:gg}
\end{eqnarray}
Performing the polarization contractions in Eq. (\ref{eq:gg}) we determine
\begin{eqnarray}
<{\cal V}_A^{ij}P^T_{ia}(\hat{\boldsymbol{k}})P^T_{jb}(\hat{\boldsymbol{k}}){\cal O}^{ab*}>&=&<{2\over d_B}\left[2(y-x_Ax_B)^2-(1-x_A^2)(1-x_B^2)\right]>\nonumber\\
&&\stackrel{t<<m_A^2,m_B^2}{\longrightarrow}<{2x_A^2\over d_Ad_B}\left(1-4x_Ax_B+x_A^2+x_B^2+x_A^2x_B^2\right)>\nonumber\\
&=&2\left(I_{20}-4I_{31}+I_{40}+I_{22}+I_{42}\right)={2\pi m_A\over \sqrt{t}}-{16\over 3}
\end{eqnarray}
so
\begin{equation}
{\rm Disc}\,{\cal M}_{grav-em}^{\gamma\gamma}(s,t)=-i{\kappa^2Z_B^2e^2\over 32\pi}{m_A\over m_B}\left({\pi m_B\over 2\sqrt{t}}-{4\over 3}\right)+B\leftrightarrow A
\end{equation}
and for the total discontinuity we determine
\begin{eqnarray}
{\rm Disc}\,{\cal M}_{grav-em}^{tot}(s,t)&=&-i{\kappa^2e^2\over 32\pi}\left[{\pi\over 2\sqrt{t}}\left(Z_B^2m_A+Z_A^2m_B\right)-{4\over 3}\left(Z_B^2{m_A\over m_B}+Z_A^2{m_B\over m_A}\right)\right.\nonumber\\
&+&\left.Z_AZ_B\left[{3\pi(m_A+m_B)\over \sqrt{t}}+6+i4\pi{m_Am_Bm_r\over p_0t}\right]\right]
\end{eqnarray}
The full amplitude is then
\begin{eqnarray}
{\cal M}^{tot}_{grav-em}(s,t)&=&G\alpha\left\{-(Z_B^2m_A+Z_A^2m_B)S-{8\over 3}\left(Z_B^2{m_A\over m_B}+Z_A^2{m_B\over m_A}\right)L\right.\nonumber\\
&+&\left.Z_AZ_B\left[-6(m_A+m_B)S+12L+i8\pi{m_Am_Bm_r\over p_0t}L\right]\right\}\nonumber\\
\qquad\label{eq:ss}
\end{eqnarray}
in agreement with the result obtained using Feynman diagrams~\cite{Ho08},\cite{Bu06},\cite{Fa07},\cite{Bo02}.  Again we observe that the amplitude contains an imaginary component arising from the second order scattering amplitude Eq. (\ref{eq:ee}), which now includes the electromagnetic potential at one vertex and the gravitational potential at the other, generating a factor of two compared to the previous cases wherein both interactions were either purely electromagnetic or gravitational.
\begin{eqnarray}
C^{(2)}_{Born}(q)&=&i\int{d^3\ell\over (2\pi)^3}{Z_AZ_Be^2\over |\boldsymbol{p}_i-\boldsymbol{\ell}|^2+\lambda^2}{i\over {p_i^2\over 2m_r}-{\ell^2\over 2m_r}+i\hat{\boldsymbol{\epsilon}}}{-Gm_Am_B\over |\boldsymbol{p}_f-\boldsymbol{\ell}|^2+\lambda^2}+Gm_Am_B\leftrightarrow Z_AZ_Be^2\nonumber\\
&&\stackrel{\lambda\rightarrow 0}{\longrightarrow}i8\pi Z_AZ_B\alpha Gm_Am_B{m_r\over p_0t}L\label{eq:ee}
\end{eqnarray}
Subtracting this term in order to determine the effective potential, we find then
\begin{eqnarray}
V_{eff}^{grav-em}(r)&=&-\int{d^3q\over (2\pi)^3}e^{-i\boldsymbol{q}\cdot\boldsymbol{r}}\left({\cal M}_{grav-em}^{tot}(q)-C^{(2)}_{Born}(q)\right)\nonumber\\
&=&G\alpha\left[{Z_A^2m_B+Z_B^2m_A\over 2r^2}+3{Z_AZ_B(m_A+m_B)\over r^2}\right.\nonumber\\
&-&\left.{4\hbar\over 3r^3}\left(Z_A^2{m_B\over m_A}+Z_B^2{m_A\over m_B}\right)+{6Z_AZ_B\hbar\over \pi r^3}\right]
\end{eqnarray}
which agrees with the forms found in \cite{Ho08},\cite{Bu06},\cite{Fa07},\cite{Bo02}.

\section{Conclusions}

Above we have calculated, using on-shell methods, the mixed electromagnetic-gravitational contribution to the scattering of two charged scalars and have shown that the result agrees with the corresponding Feynman diagram evaluation.  The latter, however, involves twenty different diagrams, each involving a four dimensional integration, while its on-shell counterpart involves only two separate contributions, both with only a two dimensional (solid angle) integration.   We have emphasized that the on-shell method inputs require only physical amplitudes so that unphysical contributions from things like ghosts are avoided.  Nowhere is this simplification more apparent than in the corresponding higher order gravitational calculation, which in the Feynman diagram case involves a myriad of indices as well as input of the three-graviton coupling,~\cite{Bo03} compared to its relatively straightforward on-shell counterpart, which is further simplified by the use of the double copy theorem~\cite{Ho17},\cite{Ch95}. The lesson is that, when applicable, on-shell methods provide a streamlined route to the evaluation of higher order nonanalytic terms and thereby to the long-range behavior.

\section{Appendix: Solid Angle Integrals}

In this appendix we give values for the various solid angle integrals $I_{mn}$ in the $t<<m_A^2,m_B^2$ limit.  We have~\cite{Fe88}
\begin{eqnarray}
I_{00}&=&-{1\over 3}+i2\pi{m_Am_Bm_r\over p_0t}+\ldots\nonumber\\
I_{11}&=&-1+\ldots\nonumber\\
I_{20}&=&{\pi m_B\over \sqrt{t}}-1+\ldots\nonumber\\
I_{02}&=&{\pi m_A\over \sqrt{t}}-1+\ldots\nonumber\\
I_{31}&=&1+\ldots\nonumber\\
I_{13}&=&1+\ldots\nonumber\\
I_{40}&=&1+\ldots\nonumber\\
I_{04}&=&1+\ldots\nonumber\\
I_{22}&=&1+\ldots\nonumber\\
I_{42}&=&{1\over 3}+\ldots\nonumber\\
I_{24}&=&{1\over 3}+\ldots\nonumber\\
I_{33}&=&{1\over 3}+\ldots\nonumber\\
I_{44}&=&{1\over 5}+\ldots\nonumber\\
\end{eqnarray}
where the ellipses indicate higher order terms in $t$.

\end{document}